%% file: Alma_es_apj.tex
\documentclass[iop]{emulateapj}

\usepackage{apjfonts,graphicx,graphics}

\usepackage{natbib}
\usepackage{latexsym}
\usepackage{amssymb}
\usepackage{longtable}
\usepackage{epsf}

\input{defn_apj}

\begin{document}

\title[Massive molecular gas flows]{Massive molecular gas flows in the Abell 1664 brightest cluster galaxy}
\author{H.~R. Russell$^1$}
\author{B.~R. McNamara$^{1,2,3}$}
\author{A.~C. Edge$^{4}$}
\author{P.~E.~J. Nulsen$^3$}
\author{R.~A. Main$^1$}
\author{A.~N. Vantyghem$^1$}
\author{F. Combes$^5$}
\author{A.~C. Fabian$^6$}
\author{N. Murray$^7$}
\author{P. Salom\'e$^5$}
\author{R.~J. Wilman$^4$}
\author{S.~A. Baum$^8$}
\author{M. Donahue$^{9}$}
\author{C.~P. O'Dea$^8$}
\author{J.~B.~R. Oonk$^{10}$}
\author{G.~R. Tremblay$^{11}$}
\author{G.~M. Voit$^{9}$}
\affil{$^1$ Department of Physics and Astronomy, University of Waterloo, Waterloo, ON N2L 3G1, Canada\\ 
    $^2$ Perimeter Institute for Theoretical Physics, Waterloo, Canada\\ 
    $^3$ Harvard-Smithsonian Center for Astrophysics, 60 Garden Street, Cambridge, MA 02138, USA\\ 
    $^4$ Department of Physics, Durham University, Durham DH1 3LE\\
    $^5$ L'Observatoire de Paris, 61 Av. de L'Observatoire, F-75 014 Paris, France\\
    $^6$ Institute of Astronomy, Madingley Road, Cambridge, CB3 0HA, UK\\
    $^7$ Canadian Institute for Theoretical Astrophysics, University of Toronto, 60 St. George St., Toronto, M5S 3H8, Ontario, Canada\\
    $^8$ School of Physics \& Astronomy, Rochester Institute of Technology, Rochester, NY 14623 USA\\
    $^{9}$ Department of Physics \& Astronomy, Michigan State University, 567 Wilson Rd., East Lansing, MI 48824 USA\\
    $^{10}$ Sterrewacht Leiden, Universiteit Leiden, P.O. Box 9513, NL-2300 RA Leiden, Netherlands\\
    $^{11}$ European Southern Observatory, Karl-Schwarzschild-Strasse 2, 85748 Garching, Germany\\}

\begin{abstract} 
We report ALMA Early Science CO(1-0) and CO(3-2) observations of the
brightest cluster galaxy (BCG) in Abell 1664.  The BCG contains
$1.1\times10^{10}\Msun$ of molecular gas divided roughly equally
between two distinct velocity systems: one from $-250$ to $+250\kmps$
centred on the BCG's systemic velocity and a high velocity system
blueshifted by $570\kmps$ with respect to the systemic velocity.  The
BCG's systemic component shows a smooth velocity gradient across the
BCG center with velocity proportional to radius suggestive of solid
body rotation about the nucleus.  However, the mass and velocity
structure are highly asymmetric and there is little star formation
coincident with a putative disk.  It may be an inflow of gas that will
settle into a disk over several $10^8\yr$.  The high velocity
system consists of two gas clumps, each $\sim2\kpc$ across, located to
the north and southeast of the nucleus.  Each has a line of sight
velocity spread of $250-300\kmps$.  The velocity of the gas in the
high velocity system tends to increase towards the BCG center and
could signify a massive high velocity flow onto the nucleus.  However,
the velocity gradient is not smooth and these structures are also
coincident with low optical-UV surface brightness regions, which could
indicate dust extinction associated with each clump.  If so, the high
velocity gas would be projected in front of the BCG and moving toward
us along the line of sight in a massive outflow most likely driven by
the AGN.  A merger origin is unlikely but cannot be ruled out.

\end{abstract}


\section{Introduction}
\label{sec:intro}


\begin{table*}
\begin{minipage}{\textwidth}
\caption{Fit parameters for the total emission line spectrum at CO(1-0) and CO(3-2).}
\begin{center}
\begin{tabular}{l c c c c c c c c c}
\hline
CO & $\nu_{\rm rest}$ & $\nu_{\rm obs}$ & $\chi^2$/dof & Gaussian & Integrated intensity\footnote{The integrated line intensities have been corrected for the primary beam response.} & Peak & FWHM\footnote{The line widths have been corrected for instrumental broadening.} & Velocity shift & Mass\footnote{The molecular gas mass was calculated from the CO(1-0) integrated intensity as described in section \ref{sec:mass}.} \\
Line & (GHz) & (GHz) & & component & (Jy{\thinspace}km/s) & (mJy) & (km/s) & (km/s) & ($10^{9}M_{\odot}$) \\
\hline
J=1-0 & 115.27 & 102.19 & 100/94 & 1 & $1.5\pm0.2$ & $8.2\pm0.8$ & $170\pm20$ & $-568\pm8$ & $5.0\pm0.7$ \\
 & & & & 2 & $1.7\pm0.3$ & $4.0\pm0.5$ & $400\pm60$ & $-60\pm20$ & $5.7\pm1.0$ \\ 
J=3-2 & 345.80 & 306.56 & 126/94 & 1 & $5.9\pm0.7$ & $29\pm2$ & $190\pm20$ & $-571\pm7$ & \\
 & & & & 2 & $6.4\pm0.9$ & $19\pm2$ & $320\pm30$ & $20\pm10$ & \\
\hline
\end{tabular}
\end{center}
\label{tab:fits}
\end{minipage}
\end{table*}

Containing molecular gas reservoirs upward of $10^9\Msun$ and in some
cases approaching $10^{11}\Msun$, brightest cluster galaxies (BCGs)
are among the most molecular gas rich early-type galaxies in the
nearby universe (\citealt{Donahue00}; \citealt{Edge01,EdgeFrayer03};
\citealt{Salome03}).  BCGs with large quantities of cool gas
reside almost exclusively at the centers of clusters whose hot
atmospheres have short central radiative cooling times ($<1\Gyr$;
\citealt{Heckman81}; \citealt{Cowie83}; \citealt{Hu85}).  In addition
to molecular gas, these BCGs are forming stars at rates of several to
tens of solar masses per year (eg. \citealt{Johnstone87};
\citealt{ODea08}).  The origin of this gas has been attributed to
either gas rich mergers or cooling from the hot atmospheres
surrounding the BCG.  However, several lines of evidence strongly
suggest that the molecular gas cooled from the hot atmosphere and
accreted onto the BCG fuelling star formation.  Although BCGs harbor
substantial amounts of molecular gas and star formation, this is
typically only a small fraction of the material that is expected to
cool out of the hot atmosphere over time (eg. \citealt{Fabian94};
\citealt{Edge01}; \citealt{Salome03}).

It is now believed that radiative cooling is regulated by
feedback from the active galactic nucleus (AGN;
eg. \citealt{PetersonFabian06}; \citealt{McNamaraNulsen07}).  This so-called
radio-mode feedback is thought to operate primarily on the hot
volume-filling atmosphere, which prevents or regulates the rate of
cooling onto the BCG and in turn the level of fuelling onto the
supermassive black hole (SMBH).  However, the degree to which
radio-mode feedback also operates on the cold molecular gas is
unclear.  Early indications that radio jets interact with the cold
atomic gas have been found in observations of ionized gas in powerful
radio galaxies (eg. \citealt{Tadhunter91}; \citealt{Veilleux02};
\citealt{Emonts05}; \citealt{Nesvadba06}) and HI absorption, which
show high speed outflows of neutral hydrogen
(eg. \citealt{Morganti98}; \citealt{Oosterloo00};
\citealt{Morganti04,Morganti05}).  However, without accurate outflow
mass estimates it is difficult to evaluate their impact on the
evolution of galaxies.


Cycle 0 ALMA observations of the Abell 1835 BCG reveal an apparent
outflow of molecular gas with a mass exceeding $10^{10}\Msun$.  Its
close association with the X-ray cavities and the
inability of the radiation and mechanical energy from supernovae to
drive such a massive flow of gas suggests that the radio AGN is
powering the outflow (McNamara et al. submitted).  These results imply
that radio mode feedback couples not only to volume-filling hot
atmospheres, but is able to drive outward the high density molecular gas
in galaxies.  Here we present new ALMA Early Science observations of
the BCG in Abell 1664 and report the discovery of massive flows of
molecular gas in the BCG.  We assume $H_0=70\kmpspMpc$, $\Omega_m=0.3$
and $\Omega_\Lambda=0.7$.  For this cosmology, $1\asec$ corresponds to
a physical scale of $2.3\kpc$ at the redshift $z=0.128$ of the BCG
(\citealt{Allen92}; \citealt{Pimbblet06}).  This redshift is based
predominantly on emission line observations and is accurate to roughly
$100\kmps$.  All errors are $1\sigma$ unless otherwise noted.



\section{Data reduction}

The BCG in Abell 1664 was observed with the band 3 and the band 7
receivers on ALMA as a Cycle 0 Early Science program (ID =
2011.0.00374.S; PI McNamara).  These data were obtained in two 25 min
observations in band 3 on March 27th and April 7th 2012 and in two 35
min observations in band 7 on March 28th 2012.  On average sixteen 12m
antennas were used in the extended configuration with baselines
out to $400\m$.  The observations used a single pointing
centred on the brightest cluster galaxy in Abell 1664.  The receiver
was tuned to cover the redshifted $^{12}$CO(1-0) line at $102.2\GHz$ in the
band 3 upper sideband and the $^{12}$CO(3-2) line at $306.7\GHz$ in the band
7 upper sideband.  Each spectral window had a bandwidth of $1.875\GHz$
and two spectral windows were set in each sideband providing a total
frequency range of $\sim{7}\GHz$.  A spectral resolution of
$0.488\MHz$ per channel was used but channels were binned together to
improve the S/N ratio.  The bright
quasar 3C\,279 was observed for bandpass calibration and observations
of Mars provided absolute flux calibration.  Observations switched
from Abell 1664 to the nearby phase calibrator J1246-257 every
$\sim10$ minutes.


The observations were calibrated using the \textsc{casa} software
(version 3.3; \citealt{McMullin07}) following the detailed processing
scripts provided by the ALMA science support team.  The
continuum-subtracted images were reconstructed using the \textsc{casa}
task \textsc{clean} assuming Briggs weighting with a robustness
parameter of 0.5 and with a simple polygon mask applied to each
channel.  This provided a synthesized beam of
$1.4\arcsec\times1.1\arcsec$ with PA $-79.1^{\circ}$ at CO(1-0) and
$0.55\arcsec\times0.39\arcsec$ with PA $-81.4^{\circ}$ at CO(3-2).
The rms noise in the line free channels was $0.5\mJy$ for $40\kmps$
channels at CO(1-0) and $1.5\mJy$ for $30\kmps$ channels at CO(3-2).
Images of the continuum emission were also produced with
\textsc{clean} by averaging channels free of any line emission.  A
central continuum source is detected in both bands, possibly partially
resolved in band 7, with flux $2.5\pm0.2\mJy$ in band 3 and
$1.1\pm0.1\mJy$ in band 7.  The mm-continuum source position coincides
with the VLA radio nucleus position (Hogan et al. in prep).  The
mm-continuum flux is also consistent, within a factor of 2, with
synchrotron emission from a flat spectrum radio core\footnote{For the
  convention $f_{\nu} \propto \nu^{-\alpha}$} with $\alpha\propto0.5$
(Hogan et al. in prep) suggesting this is the location of the low
luminosity AGN.

\section{Results}

\subsection{Spectra and integrated intensity maps}
\label{sec:specinmaps}

\begin{figure}
\centering
\includegraphics[width=0.98\columnwidth]{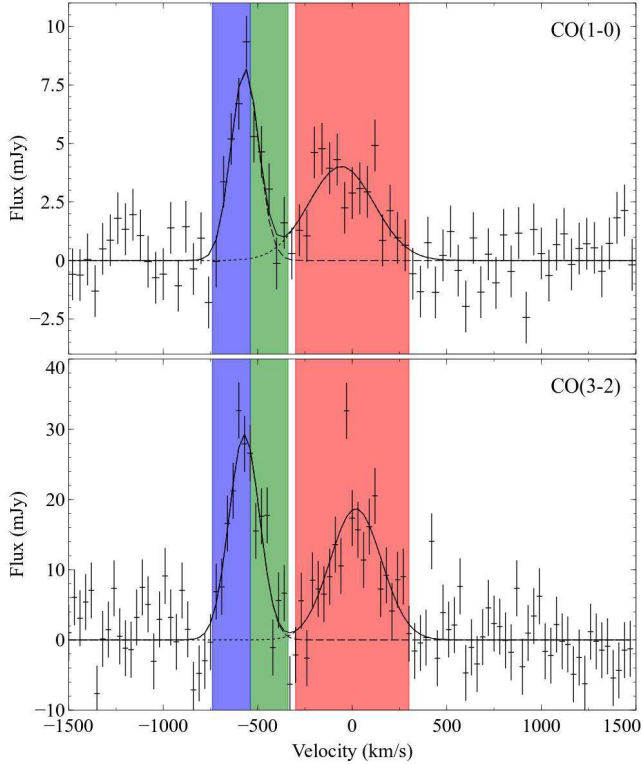}
\caption{Abell 1664 CO(1-0) (top) and CO(3-2) (bottom) total spectra for $6\times6\asec$ and $3\times3\asec$ regions, respectively.  A larger $6\times6\asec$ region at CO(3-2) produces a consistent total flux but a significantly noisier spectrum.  Two component model fits shown by solid line with individual gaussians shown by dashed and dotted lines.  The colored velocity bands correspond to the velocity ranges of the image contours in Figs. \ref{fig:CO10grid} and \ref{fig:CO32grid}.}
\label{fig:A1664spectra}
\end{figure}

We detected and imaged both CO(1-0) and CO(3-2) rotational transition
lines in the BCG of Abell 1664.  Fig. \ref{fig:A1664spectra} shows the
continuum-subtracted total spectral line profiles extracted from a
$6\times6\asec$ region at CO(1-0) and a $3\times3\asec$ region at
CO(3-2).  A larger $6\times6\asec$ region at CO(3-2) produces a consistent total flux but a significantly noisier spectrum.  Both the CO(1-0) and CO(3-2) line profiles show two distinct
components: a broad component centred on the BCG's systemic velocity and
a narrower high velocity system (HVS) blueshifted to $-570\kmps$.
These spectral profiles were each fitted with two Gaussian components using the package \textsc{mpfit} (\citealt{Markwardt09}) and the best fit
results are shown in Table \ref{tab:fits}.  Although the BCG's systemic
component is significantly broader, 
the peak flux for the HVS is
roughly a factor of two higher producing a similar integrated intensity for
each component.  The total CO(1-0) integrated intensity of
$3.2\pm0.4\Jykmps$ is roughly half of the IRAM single-dish signal
found by \citet{Edge01}.  This may be due to uncertainties in the continuum baseline subtraction, or indicate more extended emission filtered out by the interferometer or lying below our sensitivity.

\begin{figure}
\centering
\includegraphics[width=0.95\columnwidth]{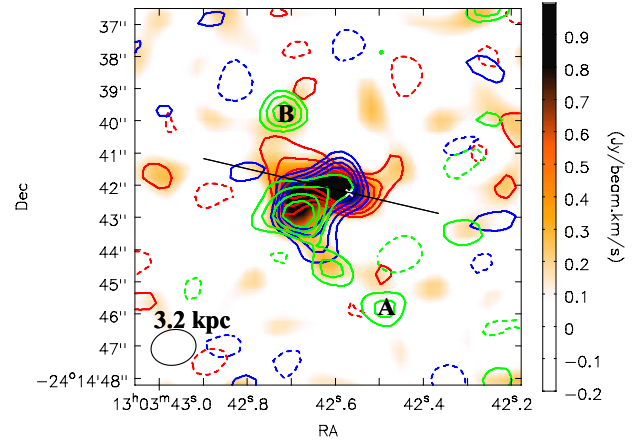}
\caption{Abell 1664 CO(1-0) integrated intensity map for velocities $-740$ to $300\kmps$ with contours for the disk component ($-300$ to $300\kmps$, red) and the high velocity system ($-740$ to $-540\kmps$, blue, and $-540$ to $-340\kmps$, green).  The contours are -2$\sigma$, 2$\sigma$, 3$\sigma$, 4$\sigma$ and 5$\sigma$, where $\sigma=0.084\Jypbmkmps$ (red), $0.047\Jypbmkmps$ (blue) and $0.046\Jypbmkmps$ (green).  The continuum nucleus is marked by a cross and the ALMA beam size is shown lower left.  The black line shows the axis of the position-velocity diagram in Fig. \ref{fig:A1664pv}.  The image has not been corrected for the primary beam.}
\label{fig:CO10grid}
\end{figure}

\begin{figure}
\centering
\includegraphics[width=0.95\columnwidth]{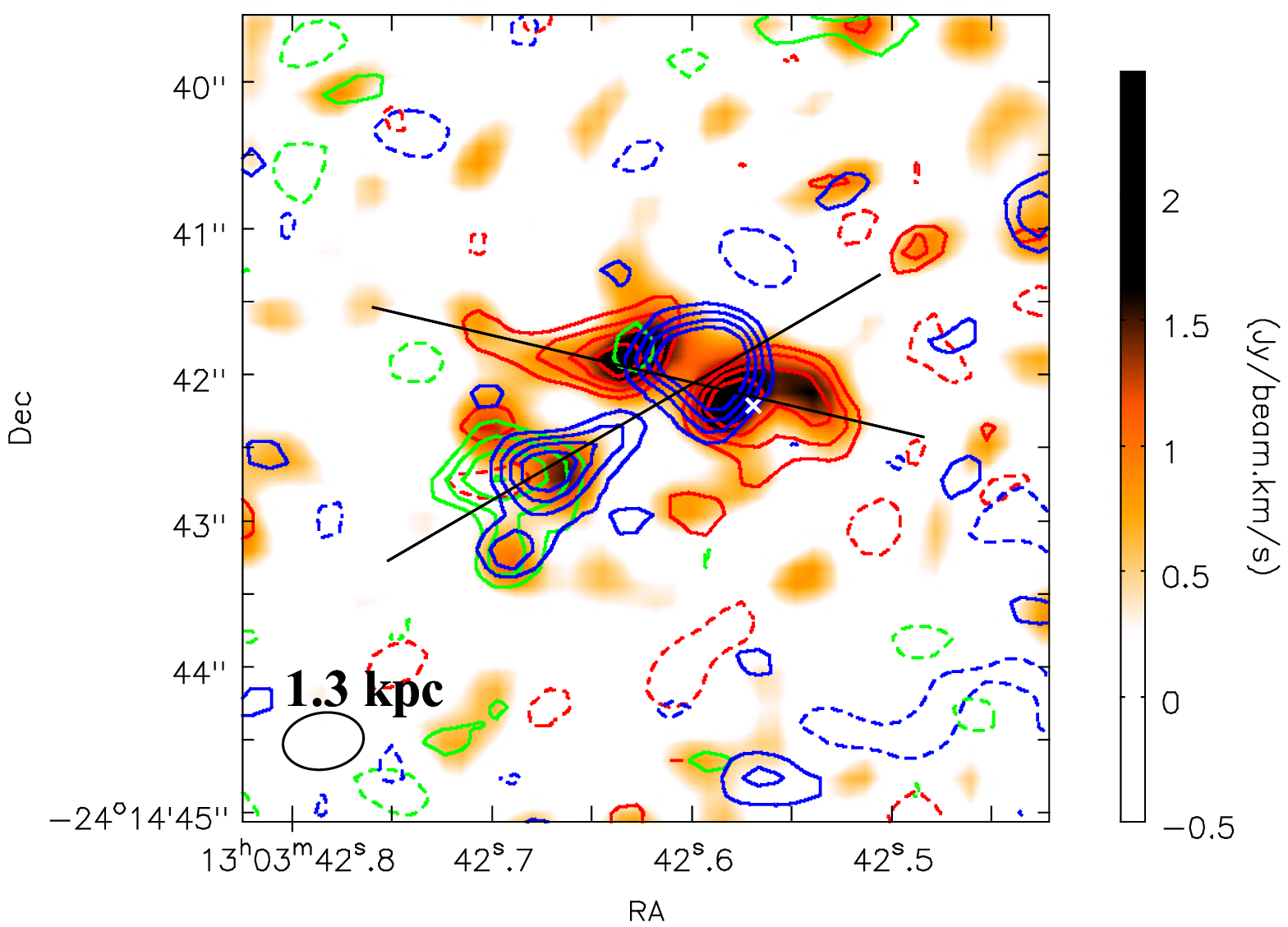}
\caption{Abell 1664 CO(3-2) integrated intensity map for velocities $-735$ to $285\kmps$ with contours for the disk component ($-285$ to $285\kmps$, red) and the high velocity system ($-735$ to $-525\kmps$, blue, and $-525$ to $-345\kmps$, green).  The contours are -2$\sigma$, 2$\sigma$, 3$\sigma$, 4$\sigma$ and 5$\sigma$, where $\sigma=0.25\Jypbmkmps$ (red), $0.15\Jypbmkmps$ (blue) and $0.13\Jypbmkmps$ (green).  Negative contours are shown by a dashed line.  The continuum nucleus is marked by a cross and the ALMA beam size is shown lower left.  The black lines show the axes of the position-velocity diagrams in Figs. \ref{fig:A1664pv} and \ref{fig:A1664CO32pvhighv}.  The image has not been corrected for the primary beam.}
\label{fig:CO32grid}
\end{figure}

Integrated intensity maps of the CO(1-0) and CO(3-2) emission are
displayed in Figs. \ref{fig:CO10grid} and \ref{fig:CO32grid}.  This
combination gives a higher resolution image of the molecular gas near
the nucleus at CO(3-2) and sensitivity over more extended scales at
CO(1-0).  The bulk of the molecular gas is concentrated in the galaxy
center within $\sim3\asec$ radius at CO(1-0) and $\sim1.5\asec$ radius
at CO(3-2).  We have produced contours from integrated intensity maps
covering the velocity range of the BCG's systemic component (red) and
the HVS (blue and green) to show their spatial extent.  Both maps show
a similar morphology with these two components separated spatially as
well as spectrally.  The kinematics of these two components broadly
match those found in H$\alpha$, Pa$\alpha$ and ro-vibrational H$_2$
IFU observations of the BCG (\citealt{Wilman06,Wilman09}).


In CO(1-0), the BCG's systemic component extends $\sim5\asec$ ($11\kpc$)
from NE to SW, covering a velocity range from $-250\kmps$ to
$+250\kmps$ and roughly peaking on the continuum position.  The line
emission from this component is highly asymmetric about the nucleus
with the greatest extent towards the NE.  There is approximately twice
as much CO(1-0) flux to the NE of the nucleus compared to the SW.  

The HVS extends $\sim4\asec$ ($9\kpc$) from NW to SE and covers a
velocity range from $-700$ to $-450\kmps$.  It appears marginally to
the N and E of the nucleus and, given the uncertainty in its position
along the line of sight through the galaxy, it is not clear if the HVS
physically interacts with the BCG's systemic component.  

The CO(3-2) map shows a broadly similar morphology to the CO(1-0) but
reveals more complex structure on smaller scales.  There is an
apparent drop in the intensity of the BCG's systemic component where it
overlaps with the HVS, possibly suggesting a physical interaction
between these components.  The SW blob of the BCG's systemic component
is roughly spatially coincident with the AGN.  The HVS may separate
into two gas blobs each contained within a projected diameter of $\sim2\kpc$.
One clump is projected immediately to the N of the nucleus, where the
deficit of emission is seen in the BCG's systemic component.  The second
HV clump appears $\sim3\kpc$ in projection to the SE and extends toward the N clump.  It could be one extended filament.

In Figs. \ref{fig:CO10grid} and \ref{fig:CO32grid}, the HVS has been
sub-divided into higher (blue) and lower (green) velocity contours to
show that the lower velocity gas in this component tends to lie at
larger distances from the nucleus.  There is a shift in the velocity
of the gas in the HVS at CO(3-2) from $-510\pm20\kmps$ in the SE blob
to $-590\pm10\kmps$ in the N clump.  The FWHM of the HVS at CO(3-2) is
$\sim230\pm30\kmps$ in the SE blob but drops to $130\pm10\kmps$ in the
clump to the N of the nucleus.  Although the separate clumps of the
HVS are spatially unresolved at CO(1-0), there is a shift in the
velocity from $-545\pm9\kmps$ to $-600\pm10\kmps$ along the HVS
filament towards the nucleus and a corresponding decrease in the FWHM
from $200\pm20\kmps$ to $130\pm30\kmps$, consistent with the CO(3-2)
results.

The gas clumps labelled A and B in Fig. \ref{fig:CO10grid} appear to
be separate structures with velocities of $-400\pm20\kmps$ and
$-370\pm40\kmps$, respectively.  Using a single Gaussian component
fitted to each spectrum, the integrated intensity was
$0.17\pm0.07\Jykmps$ for gas clump A and $0.3\pm0.1\Jykmps$ for gas
clump B.  Gas clump A is also observed at CO(3-2) with
an integrated intensity of $0.8\pm0.2\Jykmps$ but gas clump B is not
detected.


\subsection{Position-velocity diagrams}

\begin{figure*}
\begin{minipage}{\textwidth}
\centering
\includegraphics[width=0.48\columnwidth]{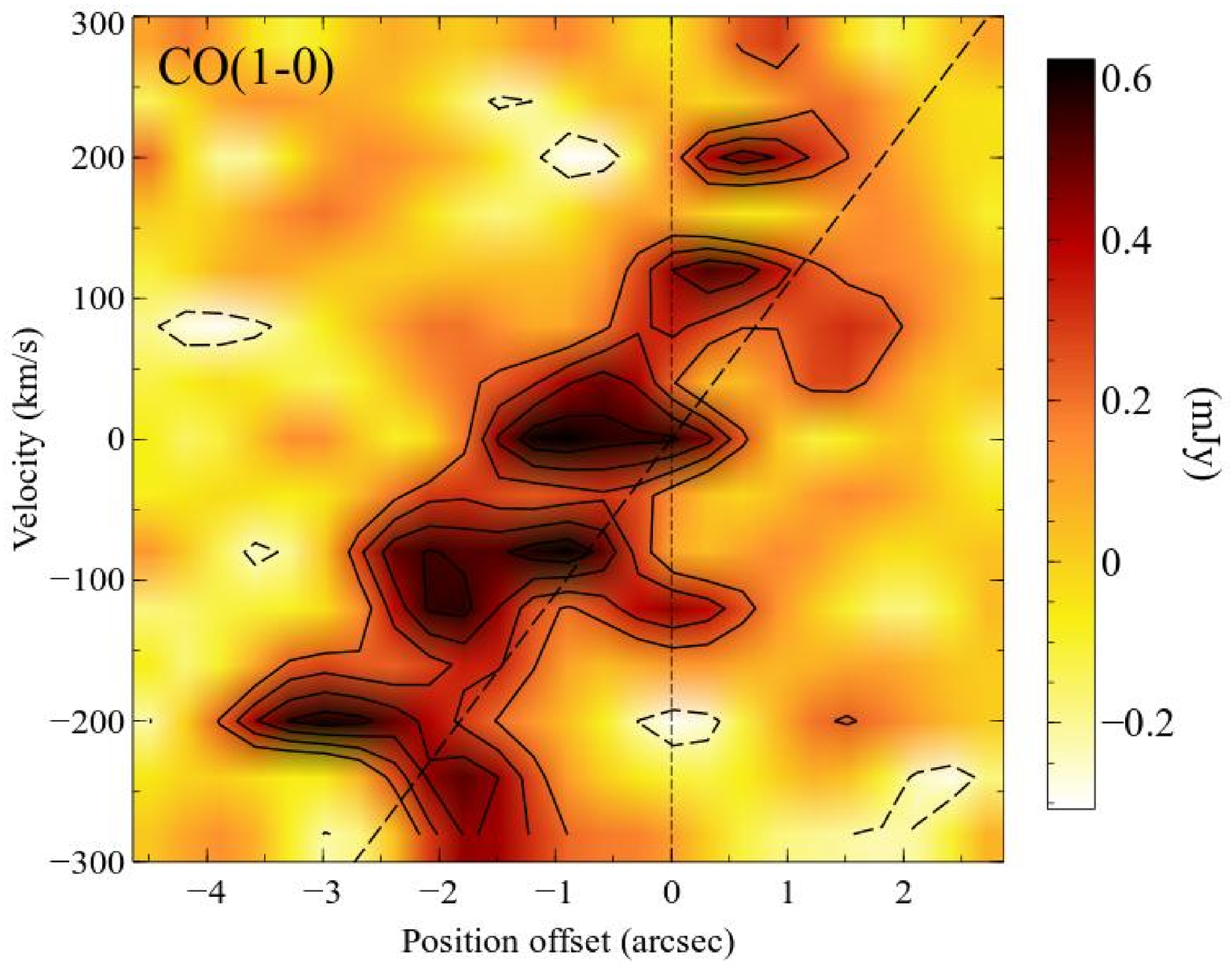}
\includegraphics[width=0.48\columnwidth]{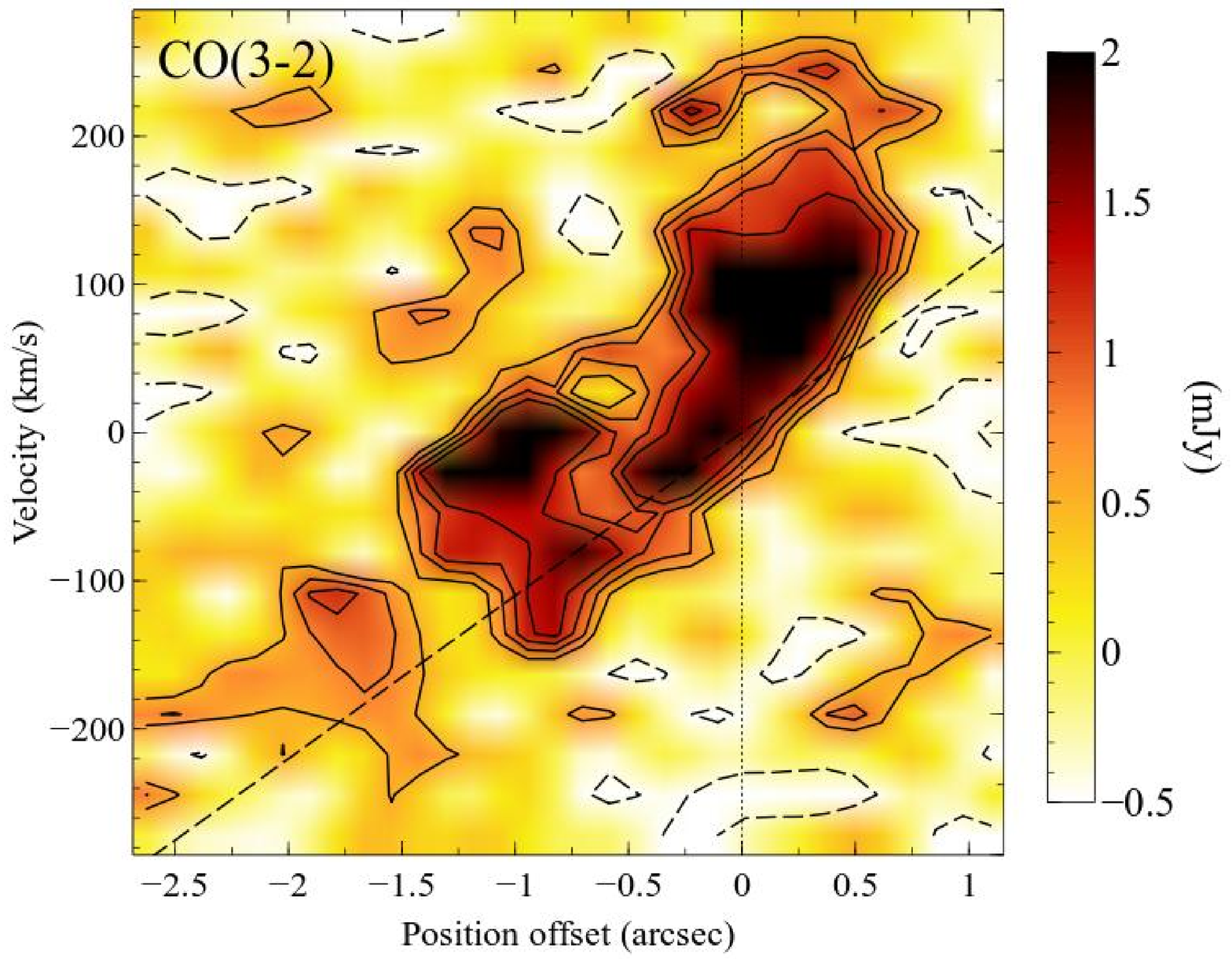}
\caption{Position-velocity diagrams for the BCG's systemic component at CO(1-0) (left) and CO(3-2) (right) each taken through a slice with position angle $77^{\circ}$.  The molecular gas in this structure extends from NE, at velocities of $-250\kmps$, to SW, at velocities of $+250\kmps$.  The contours are -2$\sigma$, 2$\sigma$, 3$\sigma$, 4$\sigma$ and 5$\sigma$.  Negative contours are shown by a dashed line.  The dotted straight line marks the continuum point source position.  The dashed straight line illustrates solid body rotation with $v{\propto}r$.} 
\label{fig:A1664pv}
\end{minipage}
\end{figure*}

\begin{figure}
\centering
\includegraphics[width=0.98\columnwidth]{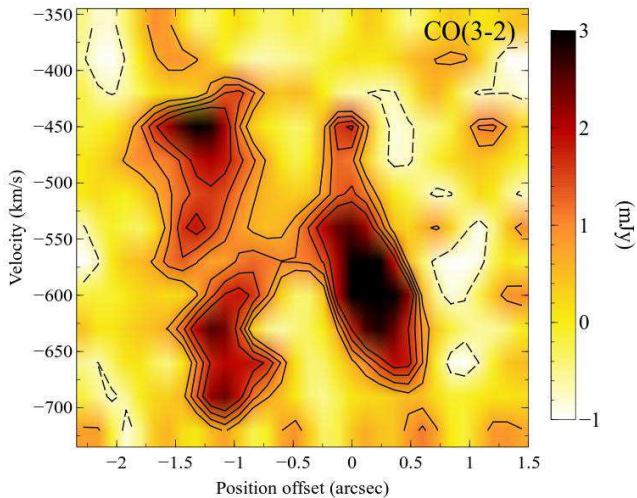}
\caption{Abell 1664 CO(3-2) position-velocity diagram for the HVS taken through a slice with position angle $120^{\circ}$.  The slice runs from SE to NW and is centred on the point where the HVS intersects the BCG's systemic component (see Fig. \ref{fig:CO32grid}).  The contours are -2$\sigma$, 2$\sigma$, 3$\sigma$, 4$\sigma$ and 5$\sigma$.  Negative contours are shown by a dashed line.}
\label{fig:A1664CO32pvhighv}
\end{figure}

Fig. \ref{fig:A1664pv} shows the position-velocity (PV) cuts along a
position angle $\rm{PA}=77^{\circ}$ (defined E from N) across the
BCG's systemic component from NE to SW at CO(1-0) and CO(3-2).  A
two-dimensional Gaussian was fitted to the integrated intensity maps
of the BCG systemic component to determine this best fit PA and this
was found to be consistent at both CO(1-0) and CO(3-2) within the
error.  The PV slices summed the line emission across the width of the
BCG's systemic component.  The CO(1-0) PV diagram shows a smooth
velocity gradient from $-250\kmps$ to $+250\kmps$, with no visible
flattening of the gradient at large radii.  The CO(3-2) emission also
traces this steady velocity gradient.  This appears consistent with
solid body rotation in the rest frame of the BCG implying a constant
mass density core over a scale of $\sim10\kpc$.  The velocity structure is similar to that of the rotating gas disk in Hydra A (eg. \citealt{Simkin79}; \citealt{Ekers83}; \citealt{Wilman05}; Hamer et al. submitted).  The AGN appears
offset from the gas dynamical center by $\sim0.5\asec$ but this may
instead be due to the uncertainty in the systemic velocity.  The
emission is clumpy and strongly asymmetric about the nucleus at both
CO(1-0) and CO(3-2) suggesting the disk may be only partially filled.



Fig. \ref{fig:A1664CO32pvhighv} shows a PV slice along the HVS from SE
to NW with best fit position angle $\rm{PA}=120^{\circ}$ at CO(3-2).
The HVS may separate into two gas blobs each with a very broad
velocity dispersion of close to $300\kmps$.  The velocity at the peak
intensity shifts from $\sim-450\kmps$ to $\sim-600\kmps$ from the SE
to NW blob but there is not a clear velocity gradient for the intervening gas as found for the
BCG's systemic component.  The PV diagram for the HVS at CO(1-0) shows a
similar velocity structure but at lower spatial resolution.


\section{Discussion}

Our CO(1-0) and CO(3-2) observations show that the BCG in Abell 1664
harbors two large gas flows projected across the galaxy center at
velocities from $-700$ to $-450\kmps$ and $-250$ to $+250\kmps$.
These molecular gas structures and their kinematics match those found
in IFU observations of the BCG covering H$\alpha$, Pa$\alpha$ and
ro-vibrational H$_2$ (\citealt{Wilman06,Wilman09}).  In the following
we calculate the molecular gas masses of these flows and consider
their possible origins including mergers, inflow settling into a disk
and outflow.

\subsection{Molecular gas mass}
\label{sec:mass}

From the integrated CO(1-0) intensities, $S_{\rm CO}\Delta v$, in Table \ref{tab:fits}, we calculated the molecular gas mass

\begin{equation}
M_{\rm mol} = 1.05\times 10^4 X_{\rm CO}\left(\frac{1}{1+z}\right) \left (\frac{S_{\rm CO}\Delta v}{\Jykmps}\right) \left(\frac{D_{\rm L}}{{\rm Mpc}}\right)^2\Msun,
\end{equation}


\noindent where $\rm X_{CO}$ is the CO to H$_2$ conversion factor
($\COtoH$), $D_{\rm L}$ is the luminosity distance and $z$ is the BCG
redshift (eg. \citealt{Bolatto13}).  Under the assumption of the
Galactic value $\rm X_{CO}=2\times 10^{20}\COtoH$, we find a molecular
gas mass for the BCG's systemic component of $M_{\rm disk} =
(5.7\pm1.0)\times 10^9\Msun$ and a mass for the high velocity system
of $M_{\rm HVS} = (5.0\pm0.7)\times 10^9\Msun$.  The total CO mass in
the BCG is then $(1.1\pm0.1)\times 10^{10}\Msun$.  The two separate gas clumps, A and B (Fig. \ref{fig:CO10grid}),  have molecular gas masses of $(6\pm2)\times10^{8}\Msun$ and $(1.0\pm0.3)\times10^{9}\Msun$, respectively.  The true value of $\rm
X_{CO}$ may depend on environmental factors such as the gas phase
metal abundance, temperature, density and dynamics.

The central gas surface density of $\sim200\Msunpsqpc$ is not unusual for a
star forming disk (eg. \citealt{Bolatto13}).  The star forming regions produce \textit{U}-band continuum emission from which \citet{Kirkpatrick09} determine a star formation rate of $\sim20\Msunpyr$.  There is no bright point source associated with the AGN.  For a radius of
$\sim3\kpc$, the surface densities of star
formation and molecular gas are ${\rm log}\Sigma_{\rm SFR}=-0.2\Msunpyrpsqkpc$ 
and ${\rm log}\mu_{\rm CO}=2.3\Msunpsqpc$,
respectively.  The Abell 1664 BCG lies with circumnuclear starbursts
and central regions of normal disks on the Schmidt-Kennicutt relation
(\citealt{Kennicutt98}) showing that despite the massive gas flows it
is similar to other normal galaxies.  The surrounding X-ray
atmosphere, from which the molecular gas presumably cooled, also has a
subsolar metallicity ($0.5-1\Zsun$, \citealt{Kirkpatrick09}).
Therefore, we argue that the Galactic $\rm X_{CO}$ value is likely to
be reasonable.  Even if $\rm X_{CO}$ lies a factor of a few below the
Galactic value, the molecular gas flows reported here would still
exceed $10^{9}\Msun$, which would not qualitatively alter our
conclusions (eg. McNamara et al. submitted).



\subsection{Mergers}

\begin{figure}
\centering
\includegraphics[width=0.48\columnwidth]{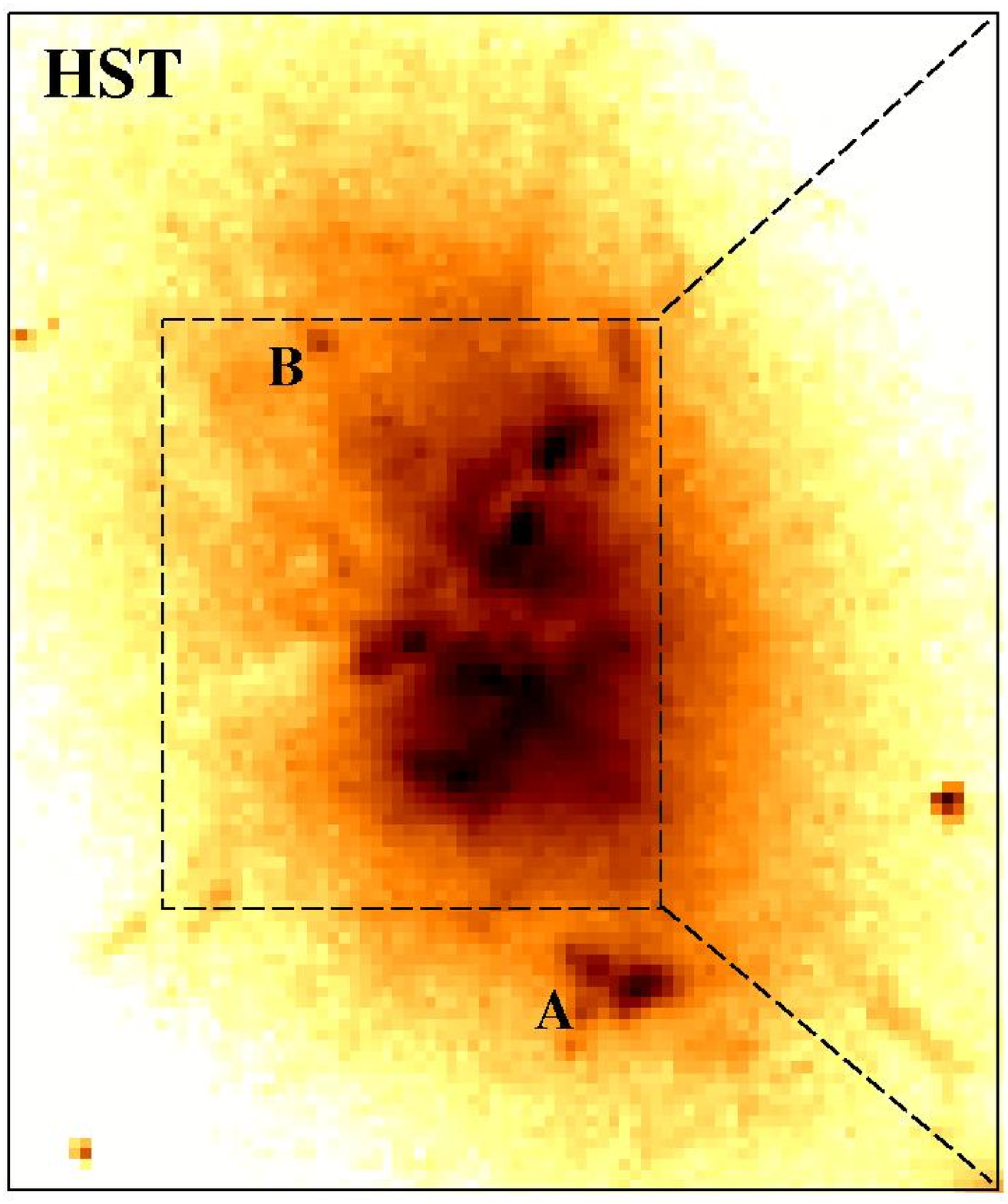}
\includegraphics[width=0.48\columnwidth]{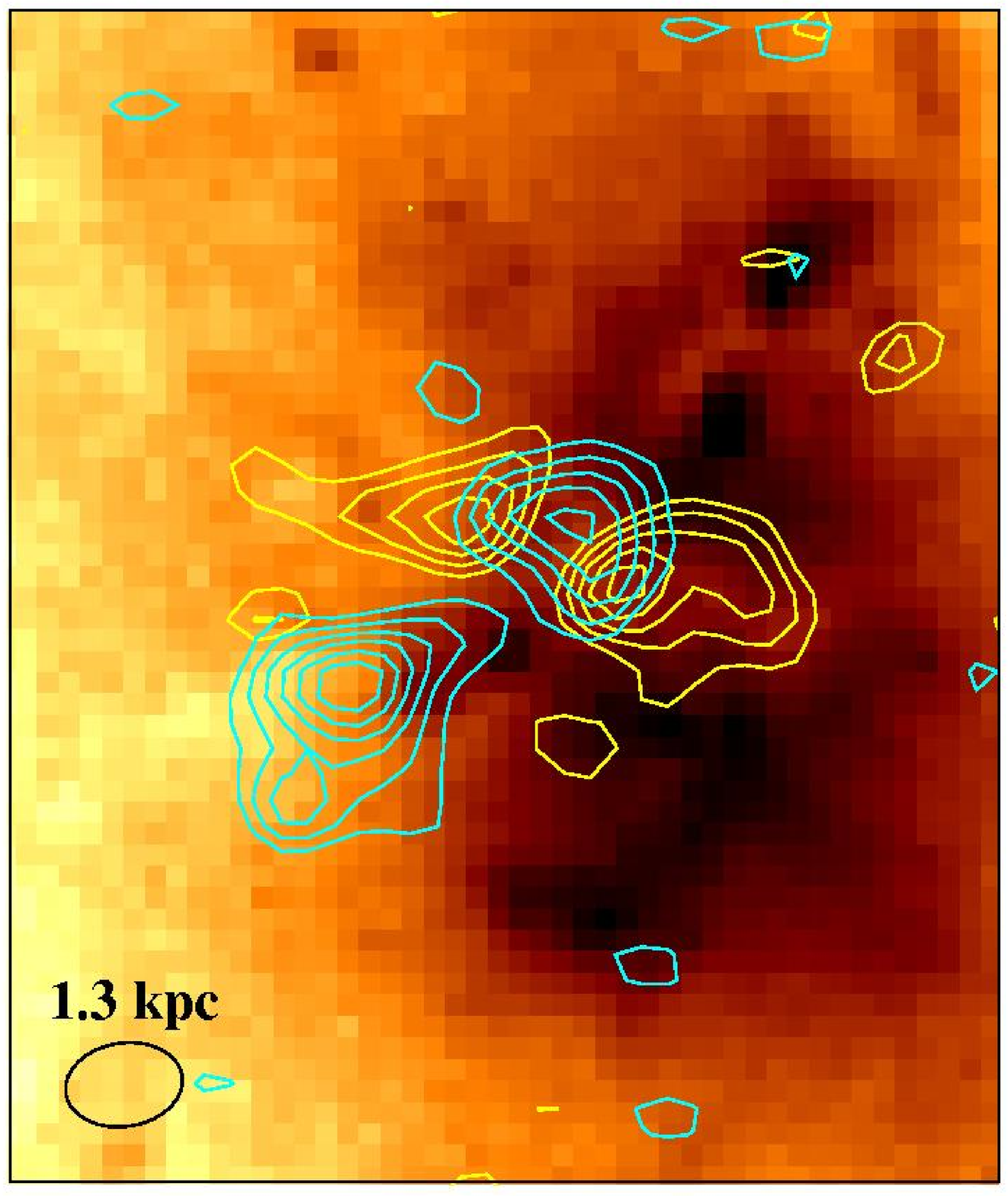}
\caption{Left: HST WFPC2 F606W optical image of the BCG in Abell 1664.  Right: Zoom in of the HST image with ALMA CO(3-2) contours representing the BCG's systemic component ($-285$ to $285\kmps$, yellow) and high velocity system ($-705$ to $-405\kmps$, cyan).  The ALMA beam size for CO(3-2) is shown to the lower left.}
\label{fig:hstabsconts}
\end{figure}

\citet{Kirkpatrick09} noted that at least four galaxies are projected
against the envelope of the BCG raising the possibility of a merger.
Although no galaxy is observed coincident with the high velocity
system, the central optical morphology is complicated by bright knots
of star formation and dust regions (Fig. \ref{fig:hstabsconts};
\citealt{ODea10}).  It is therefore difficult to rule out a collision
but there are several factors weighing against.  The BCG resides at
the center of a dense cluster atmosphere.  A galaxy plunging into the
center would likely be stripped of most of its gas before it reached
the BCG (eg. \citealt{Vollmer08}; \citealt{Kirkpatrick09};
\citealt{Jablonka13}).  Although molecular gas is more resilient to
ram pressure stripping than atomic gas, the molecular gas masses in
the BCG and the HVS are unusually large and there are few galaxies in
nearby clusters harboring this much gas (eg. \citealt{Combes07};
\citealt{Young11}).  In the context of a merger scenario, the
different velocities of the BCG's systemic component and the HVS may
require two unlikely merger events.  These interactions would
also have to be close to direct hits to the BCG center, which again
would be unlikely (eg. \citealt{Benson05}).


\subsection{Inflow}

It is more likely that such large quantities of molecular gas
originated in gas cooling from the hot cluster atmosphere
(\citealt{Edge01}; \citealt{Salome03}).  In BCGs, there are clear
associations between the shortest radiative X-ray cooling times, the
strength of optical line emission, CO and H$_2$ emission and star
formation (\citealt{Cowie83}; \citealt{Crawford99}; \citealt{Peres98};
\citealt{Rafferty08}).  The BCG in Abell 1664 is among the brightest
H$\alpha$ emitters in the \citet{Crawford99} ROSAT Brightest Cluster
Sample, the H$\alpha$ emission aligns with the soft X-ray emission
(\citealt{Allen95}) and the X-ray radiative cooling time in the center
is short ($0.4\Gyr$; \citealt{Kirkpatrick09}).  In the nearby BCG
NGC\,1275, the CO filaments are co-spatial with the coolest X-ray gas
and the outer two filaments have radial velocity gradients with larger
blueshifted velocities at smaller radii (\citealt{Salome06};
\citealt{Lim08}).  The molecular gas kinematics are consistent with
free fall in the gravitational potential of NGC\,1275, as expected if
they originated in cooling from the X-ray atmosphere.

For radial inflow, we expect velocity gradients increasing
towards the nucleus with the largest velocities towards the BCG
center.  This is more or less what is seen.  The extent and velocity
gradients across the HVS gas clumps and the BCG's systemic component can
therefore constrain their dynamics and origin
(Fig. \ref{fig:CO32grid}).  The velocity shear across the gas clumps
in the HVS is $\sim250-300\kmps$ along the line of sight
(Fig. \ref{fig:A1664CO32pvhighv}).  A similar shear in the transverse
direction could have separated the clouds on the observed scale in only $\sim10^7\yr$.
Therefore, the clumps are of order that age or the clouds are moving
nearly along the line of sight.  In
the latter case, the broad velocity shear could signify a velocity
gradient with high velocity infall onto the nucleus.


The BCG's systemic component has a clear and smooth velocity gradient
with a velocity increase of $\sim500\kmps$ over its $\sim11\kpc$
length.  The velocity profile of this component is suggestive of
rotation about the BCG center, however the mass and velocity structure
are strongly asymmetric (section \ref{sec:specinmaps}).  For a
putative rotating disk, we estimate the Toomre Q-criterion for disk
stability (\citealt{Toomre64}), $Q=v_{\rm c}v_{\rm T}/\pi Gr\Sigma$,
where the circular velocity $v_{\rm c}\sim200\kmps$, the turbulent
velocity $v_{\rm T}\sim50-100\kmps$, the disk radius $r\sim3\kpc$ and
the disk surface density $\Sigma\sim200\Msunpsqpc$.  The Q parameter
is of order one suggesting that it could be unstable and potentially
star forming.  However, Fig. \ref{fig:hstabsconts} does not show
strong star formation associated with the BCG's systemic component.
Star formation could be obscured by dust but the \textit{U}-band star
formation rate, $SFR_{\rm U}\sim20\Msunpyr$ (\citealt{Kirkpatrick09}),
and the infrared star formation rate, $SFR_{\rm IR}=15\Msunpyr$
(\citealt{ODea08}), are comparable, which is inconsistent with buried
star formation.  This could be a recent infall of material that is in
the process of settling into a disk around the nucleus.  Its orbital
time is $\sim9\times10^{7}\yr$ so a relaxed disk will take at least
several $10^8\yr$ to form.

An X-ray cooling origin for the gas inflow must explain how molecular
hydrogen can form in this environment.  It is difficult to form
molecular hydrogen rapidly in these systems by a process other than
catalytic reactions on dust grain surfaces (eg. \citealt{Ferland94}).
Hot ions in the X-ray atmosphere will destroy unshielded dust on much
shorter timescales than the radiative cooling time
(eg. \citealt{Draine79}).  Yet, dust does appear to be a ubiquitous
feature of the colder gas phases in these systems
(eg. \citealt{Egami06}; \citealt{Donahue07}; \citealt{Quillen08};
\citealt{Mittal11}).  \citealt{Ferland94} have speculated that dust
may somehow form deep within very cold gas clouds, but ab initio dust
formation by coagulation of atoms is likely to be a slow process, and
difficult in an environment where X-rays can quickly destroy small
clusters of atoms (eg. \citealt{Voit95}).  If dust-rich stellar ejecta
from the BCG and ongoing star formation can be mixed into cooling
X-ray gas before the dust grains are destroyed by sputtering, then
those seed grains may act as sites for further dust nucleation
(\citealt{Voit11}; \citealt{Panagoulia13}).  Dusty gas clouds could be
dragged out from the BCG by buoyantly rising radio bubbles and promote
subsequent molecular formation (see section \ref{sec:agnoutflow};
eg. \citealt{Ferland09}; \citealt{Salome11}).



\subsection{Outflow}

Without additional absorption line observations, it is difficult to
determine whether the gas flows lie in front or behind the BCG along
the line of sight.  Therefore, we cannot distinguish between an inflow
of gas located behind the BCG and an outflow of gas located in front
of the BCG.  Fig. \ref{fig:hstabsconts} shows that the molecular gas
clumps appear coincident with regions of lower optical-UV surface
brightness in the BCG, which could indicate either a physical
reduction in the emission or dust obscuration.  Gas clumps A and B are
associated with bright optical-UV knots.  The complexity of the
structure makes a clear interpretation difficult but it is plausible
that the dusty, molecular gas lies predominantly in front of the BCG
and the blueshifted velocities indicate an outflow.  

\subsubsection{Driving an outflow with radiation pressure or supernovae}

Radiation pressure from young stars will drive out the molecular gas
if the radiative force exceeds the gravitational force on the cloud,
$L_{\rm UV}/c > Mg$.  We estimates the gravitational acceleration as
$g=2\sigma^2/R$, where the stellar velocity dispersion for a BCG is
typically $\sim300\kmps$ (eg. \citealt{vonderLinden07}) and the
projected extent of the outflow $R=9\kpc$.  For an outflow mass
$M=5\times10^9\Msun$, this requires a luminosity in excess of
$2\times10^{46}\ergps$.  The FUV continuum emission from young stars
in the Abell 1664 BCG provides $L_{\rm UV} = 1.7\times10^{43}\ergps$
(\citealt{ODea10}), which is insufficient by three orders of
magnitude.

Mechanical winds driven by supernovae can also power molecular gas
outflows in galaxies (eg. \citealt{Veilleux05}).  Assuming
$10^{51}\erg$ per Type II supernova and a supernova production rate of
1 per $127\Msun$ (\citealt{Hernquist03}), the star formation rate in
the Abell 1664 BCG of $SFR\sim20\Msunpyr$ will generate a mechanical
power of $\sim5\times10^{42}\ergps$.  This falls short of the kinetic
power of the outflow, $E_{\rm KE}/t_{\rm
  out}\sim2\times10^{58}\erg/2\times10^{7}\yr=3\times10^{43}\ergps$,
by a factor of $\sim6$ and assumes that all of the mechanical energy
is coupled to the gas.  Little energy can be radiated.  Mechanical
winds from supernovae are therefore insufficient to accelerate the
molecular gas.

\subsubsection{Driving an outflow with a radio AGN}
\label{sec:agnoutflow}

Using ALMA Cycle 0 observations of the BCG in Abell 1835, McNamara et
al. (submitted) found a massive $>10^{10}\Msun$ outflow of molecular
gas likely driven by the radio AGN.  Active central radio sources and radio
bubbles inflated by the central AGN are ubiquitous in galaxy clusters
with short central cooling times, such as Abell 1664
(\citealt{Burns90}; \citealt{Birzan04}; \citealt{DunnFabian06}).  The
central radio source in the BCG in Abell 1664 is amorphous and fairly
weak, similar to Abell 1835.  The bulk of the energy from the radio
AGN emerges as mechanical energy, which can be estimated using scaling
relations between the jet mechanical power and the radio synchrotron
power (\citealt{Birzan08}).  Using the $1.4\GHz$ NVSS and $352\MHz$
WISH catalog radio fluxes of $36.4\mJy$ and $352\mJy$ (\citealt{Condon98}; \citealt{DeBreuck02}), respectively, \citet{Kirkpatrick09} estimated a AGN mechanical power of
$\sim6-8\times10^{43}\ergps$.  However, this has an order of magnitude
uncertainty.  The outflow velocity of the molecular gas is consistent
with the buoyancy speeds of radio bubbles.  These are typically a
significant fraction of the sound speed in the hot cluster atmosphere
(eg. \citealt{Birzan04}), which is $\sim900\kmps$ at the center of
Abell 1664.  For an outflow age of $2\times10^{7}\yr$, the energy
required to drive a $5\times10^{9}\Msun$ molecular gas outflow at a
velocity of $600\kmps$ is roughly 35\% of the AGN mechanical energy.
This is large but plausible given the order of magnitude uncertainty
on the AGN mechanical power.  

Although there is no detection of a counterpart redshifted outflow,
the asymmetry may be due to uneven mass-loading or a lopsided radio
jet in the previous outburst.  The molecular gas outflow observed in
Abell 1835 (McNamara et al. submitted) is also asymmetric, with a
higher mass and broader velocity dispersion observed in the redshifted
compared to the blueshifted component.

The putative coupling between the radio outburst and the molecular gas
is also not understood.  Simulations have shown that ram pressure
produced by high Eddington ratio jets is able to couple efficiently to
the interstellar medium and drive some of the gas out
(eg. \citealt{Wagner12}).  However, observations of Abell 1835 suggest
that the molecular gas is being lifted by the updraft in the wake of
the radio bubbles (eg. \citealt{Pope10}; McNamara et al. submitted).
It is difficult to understand how small and dense molecular gas clouds
can be uplifted with entrained hot gas.  This could potentially be
explained if the molecular gas formed in situ from hotter
$\sim0.5-1\keV$ gas rising in the bubbles' wake
(eg. \citealt{Salome08}; \citealt{Revaz08}; \citealt{Salome11}).  The
radiative cooling time of $0.5-1\keV$ gas in local pressure
equilibrium is only a few $\times10^7\yr$, which is comparable to the
rise time of the bubbles and the molecular gas.  The $0.5-1\keV$ gas
would have a density only a few times larger than the ambient hot
atmosphere but it would be several orders of magnitude less dense than
the molecular gas itself, making it much easier to lift and accelerate
to the speeds observed.  Molecular hydrogen is difficult to form
rapidly in these systems without sufficient dust and, if unshielded, dust will be
quickly sputtered by the hot X-ray gas (eg. \citealt{Draine79};
\citealt{Ferland94}).  Dust could, however, be dragged out from the
galaxy centre by rising bubbles and this may be sufficient for
molecular hydrogen formation in extended outflows
(eg. \citealt{Ferland09}).

\section{Conclusions}

Our ALMA Early Science observations of the molecular gas in the Abell
1664 BCG reveal two massive molecular gas flows each with a mass of
$M_{\rm mol}\sim5\times10^{9}\Msun$.  The component centred on the BCG
systemic velocity shows a smooth velocity gradient across the BCG
center from $-250$ to $250\kmps$ with velocity proportional to radius.
Although this is suggestive of solidy body rotation about the nucleus, the
mass and velocity structure are highly asymmetric and could 
indicate an inflow of gas in the process of forming a relaxed disk.
The high velocity system consists of two gas clumps each $\sim2\kpc$
across at a velocity of $-570\kmps$ with respect to the systemic
velocity.  Each clump has a velocity dispersion of $\sim250-300\kmps$
and there is an increase in the velocity of the gas towards the
nucleus from $-510\pm20\kmps$ to $-590\pm10\kmps$.  This velocity
gradient could signify a high velocity inflow onto the nucleus with
the broad velocity shear indicating the acceleration is along an axis
close to the line of sight.  However, the high velocity system is also
coincident with regions of low optical-UV surface brightness, which
could trace dust extinction associated with each clump.  The high
velocity gas would then be a massive outflow projected in front of the
BCG and moving toward us along the line of sight.  Based on the energy
requirements, we suggest that this outflow would most likely be driven
by the central AGN.  A merger origin for the high velocity system is
possible but we consider it unlikely.

\section*{Acknowledgements}

HRR and BRM acknowledge generous financial support from the Canadian
Space Agency Space Science Enhancement Program.  RAM and ANV
acknowledge support from the Natural Sciences and Engineering Research
Council of Canada.  ACE acknowledges support from STFC grant
ST/I001573/1.  We thank the ALMA scientific support staff members Adam
Leroy and St\'ephane Leon.  This paper makes use of the following ALMA
data: ADS/JAO.ALMA\#2011.0.00374.S. ALMA is a partnership of ESO
(representing its member states), NSF (USA) and NINS (Japan), together
with NRC (Canada) and NSC and ASIAA (Taiwan), in cooperation with the
Republic of Chile. The Joint ALMA Observatory is operated by ESO,
AUI/NRAO and NAOJ.  The National Radio Astronomy Observatory is a
facility of the National Science Foundation operated under cooperative
agreement by Associated Universities, Inc.

\lastpagefootnotes

\bibliographystyle{apj} 
\bibliography{refs.bib}

\clearpage


\end{document}

%% file: defn_apj.tex
\newcommand{\Mpc}{\rm\; Mpc}
\newcommand{\kpc}{\rm\; kpc}
\newcommand{\pc}{\rm\; pc}
\newcommand{\km}{\rm\; km}
\newcommand{\m}{\rm\; m}
\newcommand{\cm}{\rm\; cm}

\newcommand{\yr}{\rm\; yr}
\newcommand{\Gyr}{\rm\; Gyr}

\newcommand{\s}{\rm\; s}

\newcommand{\GHz}{\rm\; GHz}
\newcommand{\MHz}{\rm\; MHz}

\newcommand{\K}{\rm\; K}

\newcommand{\Msun}{\hbox{$\rm\thinspace M_{\odot}$}}

\newcommand{\Msunpsqpc}{\hbox{$\Msun\pc^{-2}\,$}}

\newcommand{\Msunpyr}{\hbox{$\Msun\yr^{-1}\,$}}
\newcommand{\Msunpyrpsqkpc}{\hbox{$\Msunpyr\kpc^{-2}\,$}}

\newcommand{\keV}{\rm\; keV}

\newcommand{\erg}{\rm\; erg}
\newcommand{\Jy}{\rm\; Jy}
\newcommand{\mJy}{\rm\; mJy}

\newcommand{\ergps}{\hbox{$\erg\s^{-1}\,$}}

\newcommand{\Jykmps}{\hbox{$\Jy\km\s^{-1}\,$}}
\newcommand{\Jypbmkmps}{\hbox{$\Jy{\rm /beam}.{\rm km}\s^{-1}\,$}}

\newcommand{\kmps}{\hbox{$\km\s^{-1}\,$}}

\newcommand{\kmpspMpc}{\hbox{$\kmps\Mpc^{-1}\,$}}

\newcommand{\Zsun}{\hbox{$\thinspace \mathrm{Z}_{\odot}$}}

\newcommand{\asec}{\rm\; arcsec}

\newcommand{\psqcm}{\hbox{$\cm^{-2}\,$}}

\newcommand{\COtoH}{\hbox{$\psqcm(\K\kmps)^{-1}$}}

